\input{aipcheck}

\documentclass[
    ,final            
  ]
  {aipproc}

\layoutstyle{6x9}

\begin{document}

\title{Stellar Age Estimation from ~3 Myr to ~3 Gyr }

  

\author{Lynne Hillenbrand}{
  address={California Institute of Technology}
}

\author{Eric Mamajek}{
  address={Harvard-Smithsonian Center for Astrophysics}
}

\author{John Stauffer}{
  address={Spitzer Science Center}
}

\author{David Soderblom}{
  address={Space Telescope Science Institute}
}

\author{John Carpenter}{
  address={California Institute of Technology}
}

\author{Michael Meyer}{
  address={The University of Arizona}
}

\begin{abstract}
 We present recent progress on quantitative estimation of stellar ages using indicators such as theoretical evolutionary tracks, rotation, rotation-driven chromospheric and coronal activity, and lithium depletion. Our focus is on roughly solar-mass and solar-metallicity stars younger than the Sun.  We attempt to characterize the systematic and random error sources and then derive ``best'' ages along with the dispersion in age arising among the various age estimation methods.  Our main application of these techniques is to the evolution of debris disks. 
 \end{abstract}

\maketitle


\section{Introduction and Age Dating Methods}
 Fundamental stellar properties such as mass, radius, and rotation speed are derivable -- for a limited set of stars at present -- through basic observables such as orbital motion, eclipses, and period measurements.  In contrast, stellar ages have no firm basis or anchor other than in the case of the Sun for which radiometric dating of primitive solar system materials is possible and gives the solar age to within a few Myr. Yet, stellar ages are critical to establish for investigations such as the time scales involved in the formation and long term evolution of planetary systems.  

Recent progress has been made on the quantitative estimation of ages for stars of roughly solar-mass and solar-metallicity that are younger than the Sun.
Methods for estimating stellar ages include:
\begin{itemize}
\item
{\it Hertzsprung-Russell diagram.}  For stars younger than the Sun, this purely theoretical age estimator is useful only with those $<$20-30 Myr old, i.e. in the pre-main sequence phase. Track-dependent systematics can be large.
\item
{\it Chromospheric activity.}  R'$_{HK}$ = (L$_{HK}$/L$_{bol}$) ranges over $\sim$1 dex, from 10$^{-4}$ to 10$^{-5}$ with saturation above -4.35 dex where increasing rotation leads to relatively little increase in CaII H\&K line emission activity.
\item
{\it Coronal activity.}  R'$_{xray}$ = (L$_{xray}$/L$_{bol}$) ranges over $\sim$4 dex, from 10$^{-3}$ to 10$^{-7}$ with saturation in soft x-ray activity above -4.0 dex.
\item
 {\it Rotation.} The chromospheric and coronal activity are rotation driven. Periods range from $<$1 to $>$40 days; projected velocities range from $>$150 to $<$2 km/s. Stellar rotation has maximum spread around the age of Alpha Per ($\sim$80 Myr), likely related to dispersion in the previous history of interaction between the star and its circumstellar disk. Standard main sequence spin-down obtains thereafter.
\item
 {\it Lithium.} Depletion of light elements is related both to standard stellar evolution and to rotation.  Li I $\lambda$6707 is required for youth but generally has a factor of two empirical spread at constant age and mass.
\end{itemize}

\section{Calibration of Activity-Rotation-Lithium to Age}


\begin{figure}
  \includegraphics[height=.33\textheight]{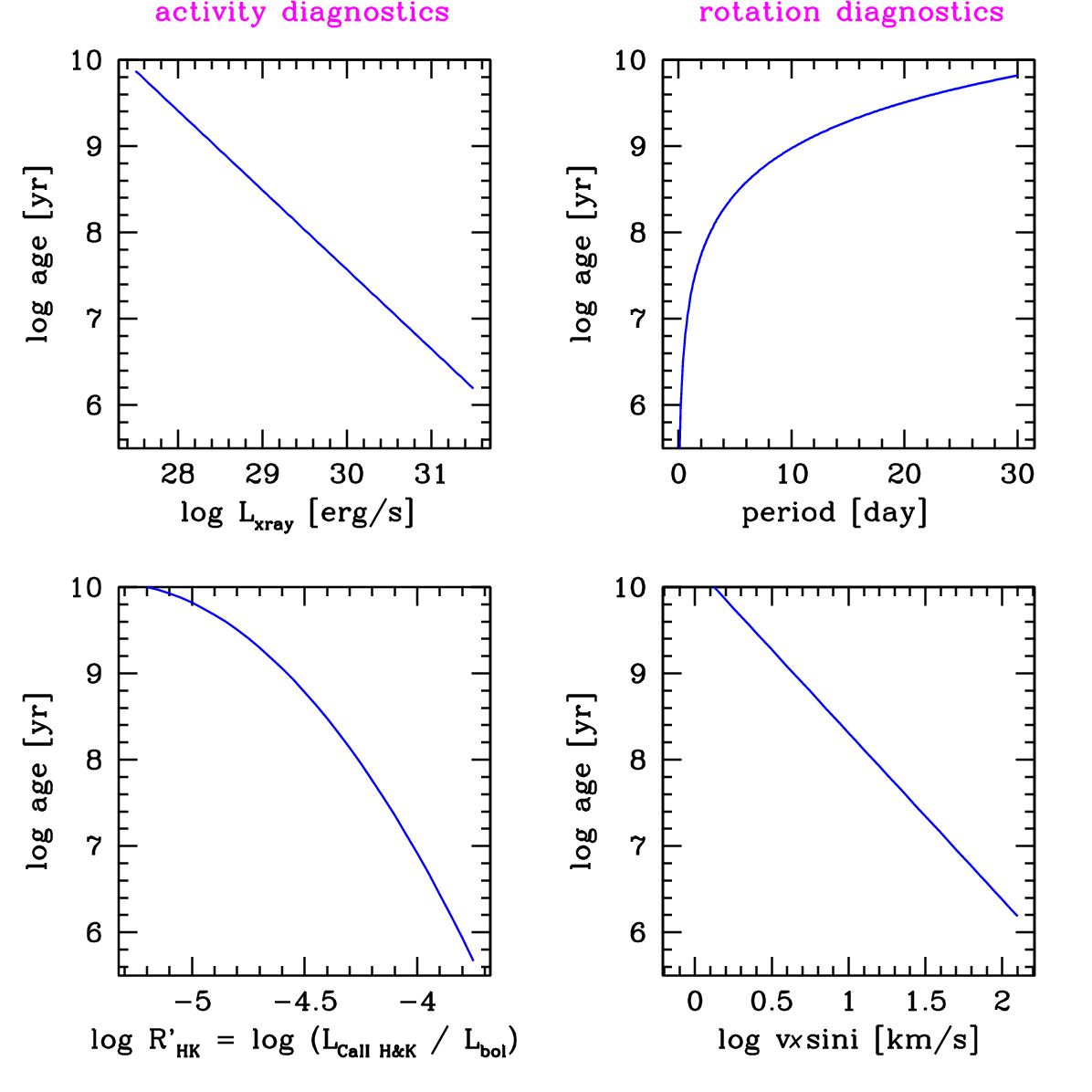}
  \caption{
 State-of-the-art calibrations between observed diagnostics and stellar age, 
shown for solar color stars.   Color-dependent equations have been derived 
using members of clusters of "known" age that are themselves dated using 
theoretically validated methods such as the high mass
upper main sequence turnoff 
and the low mass lithium depletion boundary. }
\end{figure}

Notably, the behavior of each of the above empirical quantities with age is mass dependent!
Our calibrations to age are established using critically scrutinized samples of 
open cluster stars and we utilize visual binaries for consistency checks.  
New activity-age and rotation-age calibrations appear in Figure 1.
We have updated existing chromsopheric activity-age relations from Soderblom et al. (1991), Donahue (1993), and Lachaume et al. (1999), populating the high activity end for the first time and using modern open cluster ages.  We have also updated the Sterzik \& Schmitt (1997) coronal-chromospheric correlations and tied both activity indicators through the Rossby number to rotation.  We have corrected the Barnes (2007) gyrochronology relations to match the Pleiades, Hyades, and Sun, and show consistency with binary pairs; we have also derived empirical vsini relations.  See Mamajek (this proceedings) for equations and demonstration of the 0.2 dex or better accuracy in our new activity-rotation-age calibrations for stars aged between the Hyades (600 Myr) and the Sun (4560 Myr). At younger ages, accuracy is only 0.3-1.0 dex and the Li I and H-R diagram methods may be more accurate.  
We also consider the lithium depletion trends in young clusters and devise a probabilistic age estimator that accounts for the observed dispersion.

\section{Field Star Age Derivation}

\begin{figure}
  \includegraphics[height=.6\textheight,angle=-90]{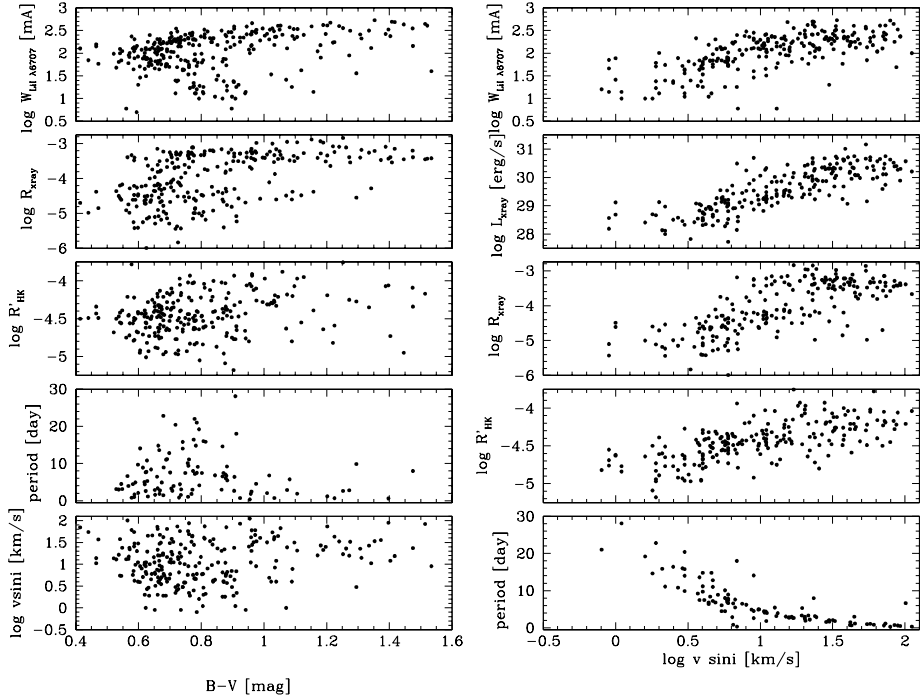}
  \caption{Range in empirical age diagnostics with B-V 
  (left panels) and vsin$i$ (right panels).}
\end{figure}

We have assembled a database on empirical age indicators
relevant to the Spitzer/Legacy Program 
FEPS (Meyer et al. 2006), deriving from both the literature 
and data newly obtained for this purpose.  As shown in Figure 2,
among our field star sample there is a large spread at any given color in every age diagnostic, reflecting the spread in age as well as astrophysical dispersion.
The empirical age indicators are well correlated with one another and their dispersion highlights the astrophysical spread plus the color effects.  X-ray and probable lithium saturation effects are apparent at high activity levels.

\begin{figure}
  \includegraphics[height=.30\textheight]{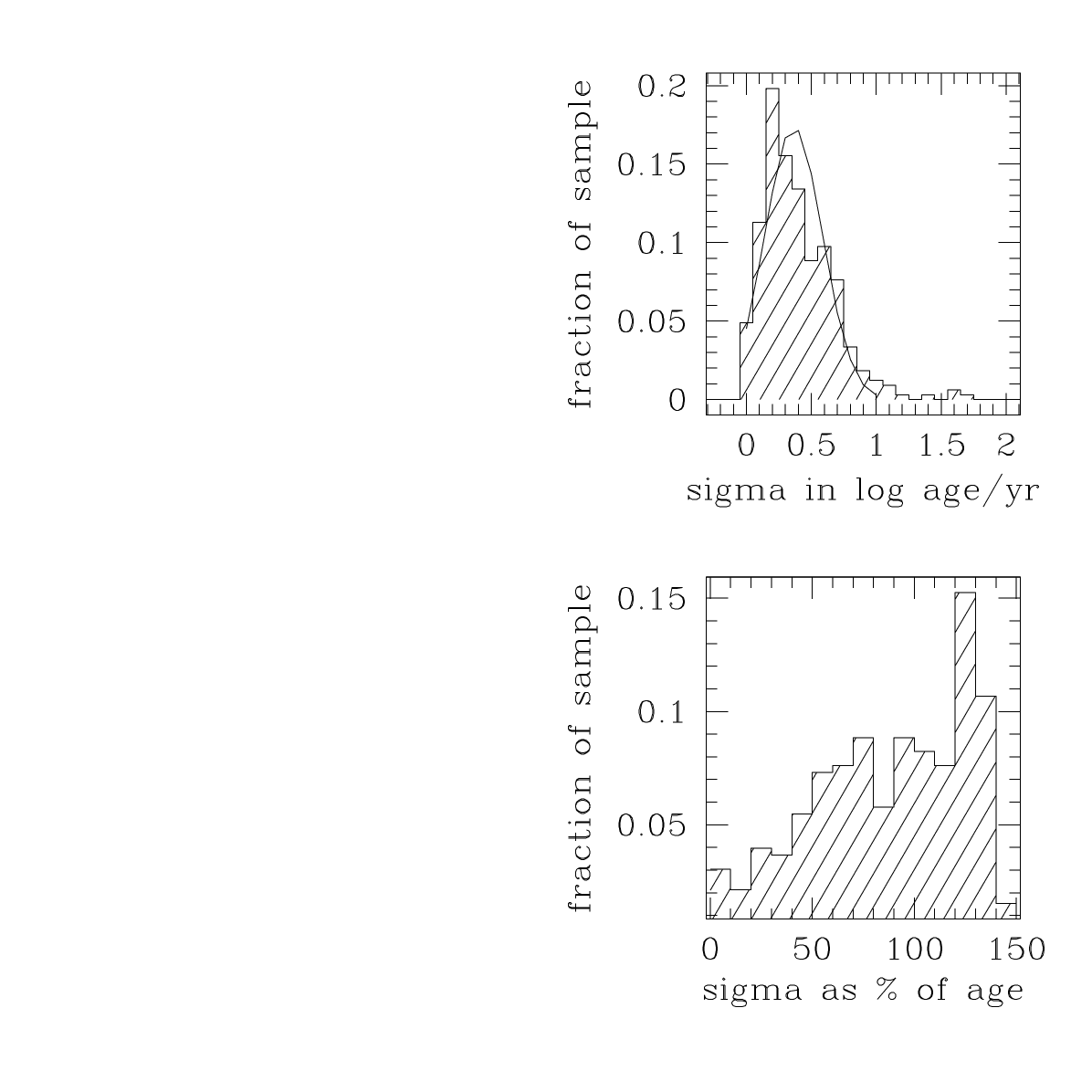}
  \includegraphics[height=.30\textheight]{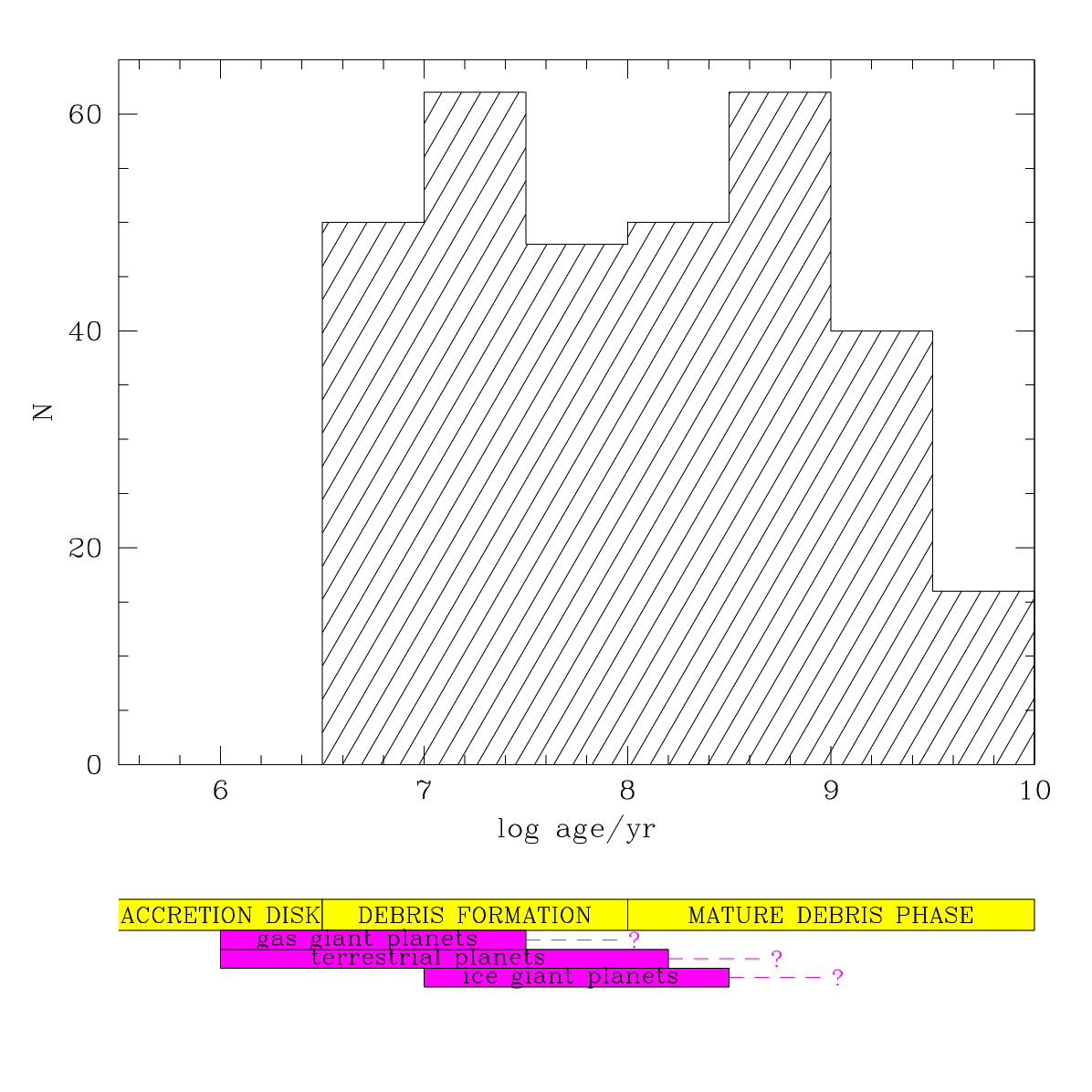}
  \caption{On left is the dispersion in ages 
  derived from the different diagnostics,
  shown both in dex and as a percentage of mean age.  The Gaussian
  is drawn with mean 0.35 dex and width 0.18 dex, 
  though the data peak closer to 0.2 dex.
  On right is the final age distribution for stars on the FEPS program.
  }
\end{figure}

In Figure 3 we display the dispersion in age resulting from application 
of the ensemble of age estimation techniques to each star, as well as
the mean ages.  The distribution of dispersions indicates mean value
0.35 dex though the data have a sharper peak closer to only 0.2 dex.
The dispersion is higher for younger stars, and the tail 
of large dispersion values is indeed dominated by younger stars.  
According to these distributions, age errors of 25-150\% should be
those typically quoted for the ages of stars younger than the Sun.

From our study of stellar age dating techniques in the 3 Myr to 3 Gyr age range:
\begin{itemize}
\item

Cluster membership most securely establishes a stellar age. 
 
\item
The H-R diagram can and should be used at <20-30 Myr, with confirmation of youth coming from other activity/lithium diagnostics.
 
\item
Lithium buffered by activity/rotation diagnostics can be used ~30-200 Myr.
 
\item
Activity/rotation is most useful for slow rotators, age >200 Myr.
\end{itemize}
Significant age ambiguity remains for field stars younger 
than the Sun!  Uncertainties in calibration ($<$0.2 dex) 
combined with empirical spreads in the age indicators 
(0.4 $\pm$ 0.3 dex) suggest substantial age uncertainty
between $\sim$30 Myr and $\sim$3 Gyr.



\section{Application to Debris Disk Evolution}

Results from e.g.  Meyer et al. (2008),
Hillenbrand et al. (2008), and Carpenter et al. (2009), 
on the evolution of dust signatures 
from debris disks rely on knowledge of stellar ages.  From application
of the techniques discussed here to our Spitzer sample, 
we infer that $\sim$300 Myr marks a transitional time in debris disk 
evolution as evidenced by breaks in dust detection frequency and luminosity.  
Older stars have weaker disks, roughly consistent with steady state 
collisional evolution models.  At least 15\% of solar type stars form 
debris disks but with inner cleared regions several tens of AU in size 
and surface densities possibly several times lower than that inferred 
for the young Kuiper Belt, though this last result is heavily model dependent.





\bibliographystyle{aipproc}   


\begin{thebibliography}{ }
\bibitem{junk1}
Barnes, S.A., 2007, ApJ, 669, 1167
\bibitem{junk2}
Carpenter, J.M., et al.\ 2009, ApJ, in press
\bibitem{junk3}
Donahue, R.A.\ 1993, Ph.D.~Thesis, New Mexico State University
\bibitem{junk4}
Hillenbrand, L.A., et al. 2008, ApJ, 677, 630
\bibitem{junk5}
Lachaume, R., Dominik, C., Lanz, T., \& Habing, H.J.\ 1999, A\&A, 348, 897
\bibitem{junk6}
Meyer, M.R., et al.\ 2008, ApJL, 673, 181
\bibitem{junk7}
Meyer, M.R., et al.\ 2006, PASP, 118, 1690
\bibitem{junk8}
Soderblom, D.R., Duncan, D.K., \& Johnson, D.R.H.\ 1991, ApJ, 375, 722
\bibitem{junk9}
Sterzik, M.F., \& Schmitt, J.H.M.M.\ 1997, AJ, 114, 1673

\end{thebibliography}

\IfFileExists{\jobname.bbl}{}
 {\typeout{}
  \typeout{******************************************}
  \typeout{** Please run "bibtex \jobname" to optain}
  \typeout{** the bibliography and then re-run LaTeX}
  \typeout{** twice to fix the references!}
  \typeout{******************************************}
  \typeout{}
 }



\end{document}